\documentclass{aa}

\usepackage{graphicx}
\usepackage{txfonts}
\usepackage{color}
\begin{document}

   \title{Planetesimal formation at the gas pressure bump following a migrating planet}

   \subtitle{I. Basic characteristics of the new formation model}

   \author{Y. Shibaike\inst{1}
          \and
          Y. Alibert\inst{1}
          }

   \institute{Physikalisches Institut \& NCCR PlanetS, Universitaet Bern, CH-3012 Bern, Switzerland\\
              \email{yuhito.shibaike@space.unibe.ch}
             }

   \date{Received MM DD, 2020; accepted MM DD, 2020}

 
  \abstract
   {To avoid known difficulties in planetesimal formation such as the drift or fragmentation barriers, many scenarios have been proposed. However, in these scenarios, planetesimals form in general only at some specific locations in protoplanetary discs. On the other hand, it is generally assumed in planet formation models and population synthesis models, that planetesimals are broadly distributed in the protoplanetary disc.}
   {Here we propose a new scenario in which planetesimals can form in broad areas of the discs. Planetesimals form at the gas pressure bump formed by a first-generation planet (e.g. formed by pebble accretion) and the formation region spreads inward in the disc as the planet migrates.}
   {We use a simple 1D Lagrangian particle model to calculate the radial distribution of pebbles in the gas disc perturbed by a migrating embedded planet. We consider that planetesimals form by streaming instability at the points where the pebble-to-gas density ratio on the mid-plane becomes larger than unity. In this work, we fix the Stokes number of pebbles and the mass of the planet to study the basic characteristics of this new scenario. We also study the effect of some key parameters like the ones of the gas disc model, the pebble mass flux, the migration speed of the planet, and the strength of turbulence.}
   {We find that planetesimals form in wide areas of the discs provided the flux of pebbles is typical and the turbulence is not too strong. The planetesimal surface density depends on the pebble mass flux and the migration speed of the planet. The total mass of the planetesimals and the orbital position of the formation area depend strongly on the pebble mass flux. We also find that the profile of the planetesimal surface density and its slope can be estimated by very simple equations.}
   {We show that our new scenario can explain the formation of planetesimals in broad areas. The simple estimates we provide for the planetesimal surface density profile can be used as initial conditions for population synthesis models.}


   \maketitle
%

\section{Introduction}
\label{introduction}
Explaining theoretically the formation of planetesimals is known to suffer from many difficulties. One of the most significant issues in the classical formation scenario, collisional growth of dust particles, is the drift of the particles. Particles indeed suffer head wind from the gas part of the disc, this latter rotating with a sub-Kepler speed due to the gas pressure gradient, resulting in losing their angular momentum and drifting toward the central star \citep[e.g.][]{whi72}. To avoid this ``drift barrier'', a lot of scenarios have been proposed. For example, one may invoke the evolution of the internal density of the particles \citep[e.g.][]{oku12}, or consider the formation at the local structures in the gas discs; the water snow line \citep[e.g.][]{sai11}, the boundaries of the dead zone \citep[e.g.][]{bra08},  vortices in the gas \citep[e.g.][]{bar95}, or  gas pressure bumps formed by magnetically driven disc wind or by embedded planets
\citep[e.g.][]{tak18,paa04}. Fragmentation of the particles due to their fast mutual collision speed is another problem especially for rocky planetesimal formation \citep[e.g.][]{blu93}. Bouncing and charging of the particles also prevent their collisional growth\citep[e.g.][]{zso10,oku09}. Therefore, one in general need to have some sort of instabilities at some specific locations where the particles accumulate, to  efficiently  form planetesimals \citep[e.g.][]{sek98,you05}.

However, with these previous scenarios, planetesimals can form only at some specific places in the discs. On the other hand, existing population synthesis models generally assume that the radial distribution of planetesimals is much broader and continuous (no gap exists in the distribution). \citet{len19} also pointed out this issue and showed that planetesimals can form in wide regions if the instabilities can occur in the entire regions of the discs. They argued that this assumption is supported by the observations; the ring structures which can accumulate solid particles exist at several places in the discs. However, if the structures are fixed at the places, subsequent planetesimal formation occurs only there and the discs have regions where the planetesimals do not exist. There are indeed some previous works which investigated the planetesimal formation at the gas pressure bumps formed by embedded planets but the planetesimals in their results are concentrated at several locations because of their stopped or very limited planetary migration \citep[e.g.][]{lyr09,ayl12a,cha13,sta19,eri20}.

In this paper, we propose a new scenario for planetesimal formation which can operate in broad areas of the discs. Once a first-generation planet forms in the outer regions of a protoplanetary disc (by any mechanism e.g. pebble accretion), it rapidly forms a pressure bump beyond its orbit. Recent observations of a young disc ($<0.5~{\rm Myr}$ old) with gap structures actually suggest that such first-generation planets can exist early \citep{seg20}. Drifting pebbles then accumulate there and planetesimals may form by streaming instability due to the high local density of pebbles. The place where the planetesimals form then moves inward following the planet inward migration, resulting in the formation of planetesimals in broad regions of the discs.

In this paper (Paper I), we expose the basic features of this new scenario and show the results with simple assumptions and calculations. In a future paper (hereafter Paper II), we will show the cases with more realistic situations including growth of pebbles and planets.
  
\section{Methods}
\subsection{Gas disc models}
\label{gas}
In this work, we assume that the underlying gas disc structure does not depend on time. The unperturbed (non perturbed by the planet) gas surface density is
\begin{equation}
\Sigma_{\rm g,unp}=\Sigma_{\rm g,0}\left(\dfrac{r}{\rm au}\right)^{-p},
\label{gasg}
\end{equation}
and the disc temperature (on the mid-plane) is
\begin{equation}
T=T_{\rm 0}\left(\dfrac{r}{\rm au}\right)^{-q},
\label{temperature}
\end{equation}
where $r$ is the distance from the central star. Here, we will consider three different models of protoplanetary discs: discs A, A' and B. The parameters $\Sigma_{\rm g,0}$, $T_{\rm 0}$, $p$, and $q$ of the three discs are given in Table \ref{tab:discs}. Disc A is consistent with the observations of protoplanetary discs under the assumption that the dust-to-gas surface density ratio is uniform through the entire discs \citep{and10}. Disc A' is a ten times heavier and $10^{1/4}$ times hotter disc compared to Disc A with the same temperature structure. Disc B is the minimum-mass solar nebular \citep{hay81}.

\begin{table}[h]
\caption{Parameters of the discs}
\label{tab:discs}
\centering
\begin{tabular}{lllll}
\hline
Disc & $\Sigma_{\rm g,0}$ & $T_{\rm 0}$ & $p$ & $q$ \\
\hline\hline
A & 500  & 280 & 1 & 0.5 \\
A' & 5000  & 500 & 1 & 0.5 \\
B & 1700 & 280 & 1.5 & 0.5 \\
\hline
\end{tabular}
\end{table}

\subsection {An embedded planet}
\label{planet}
We assume a single planet with fixed mass of $20~M_{\rm E}$ at $r=30~{\rm au}$ exists at the beginning of the calculation. The embedded planet influences the gas disc and changes the gas structure around while migrating at the same time. 

\subsubsection{Gap structure}
\label{gap}
The embedded planet excites density waves by gravitational interaction with the gas disc. The waves are then damped due to the disc viscosity or non-linear effects, resulting in the deposition of the angular momentum in the disc. Here, we consider the following approximate formulas which are functions of $r$ to describe the gap structure of the gas disc around the planet. Although the functions are derived under a number of assumptions (steady state, no mass flow toward the central star, Keplerian rotation, instantaneous damping excited density waves, no migration of the planet - see \citet{tan07,kan15a}), we use the formulas in this work for simplicity (see also Appendix \ref{etazero}). We assume that the local gas surface density around the planet is described as,
\begin{equation}
\Sigma_{\rm g,local}=\Sigma_{\rm g,unp}\max(s_{\rm K}, s_{\rm min}),
\label{gasl}
\end{equation}
where the factor $s_{\rm K}$ is $s_{\rm K}=\max(s_{\rm Kepler}, s_{\rm Rayleigh})$. The factor $s_{\rm Kepler}$ is then,
\begin{equation}
s_{\rm Kepler}=\begin{cases}
\exp\left(-\dfrac{C}{9|x|^{3}K}\right) & (|x|>\Delta) \\
\exp\left(-\dfrac{C}{9\Delta^{3}K}\right) & (|x|\leq\Delta),
\end{cases}
\label{sKepler}
\end{equation}
where $C=0.798$, $\Delta=1.3$, and $x=(r-r_{\rm pl})/H_{\rm g,pl}$. Here, the subscript ``pl'' indicates the value at the location of the planet. The gas scale height is $H_{\rm g}=c_{\rm s}/\Omega_{\rm K}$, where the sound speed is $c_{\rm s}=\sqrt{k_{\rm B}T/m_{\rm g}}$ and the Kepler frequency is $\Omega_{\rm K}=\sqrt{GM_{*}/r^{3}}$. The Boltzmann constant is $k_{\rm B}$, the gravitational constant is $G$, and the mean molecular mass is $m_{\rm g}=3.9\times10^{-24}~{\rm g}$. The mass of the central star, $M_{*}$, is one solar mass. To keep the dynamical stability, the Rayleigh stability condition \citep[e.g.][]{cha61}, which could be violated near the planet, must be satisfied in the whole region. The factor $s_{\rm Rayleigh}$ is then, from the continuity of the surface density considered,
\begin{equation}
s_{\rm Rayleigh}=\begin{cases}
\exp\left(-\dfrac{5}{6}x_{\rm m}^{2}+\dfrac{5}{4}x_{\rm m}|x|-\dfrac{1}{2}x^{2}\right) & (|x|>\Delta) \\
\exp\left(-\dfrac{5}{6}x_{\rm m}^{2}+\dfrac{5}{4}x_{\rm m}\Delta-\dfrac{1}{2}\Delta^{2}\right) & (|x|\leq\Delta),
\end{cases}
\label{sRayleigh}
\end{equation}
where $x_{\rm m}=\{(4/3)CK\}^{1/5}$ is the outer edge of the marginal Rayleigh stable region (i.e., $|x|<x_{\rm m}$). The factor $K$ is given by
\begin{equation}
K=\left(\dfrac{M_{\rm pl}}{M_{*}}\right)^{2}\left(\dfrac{r_{\rm pl}}{H_{\rm g,pl}}\right)^{5}\alpha^{-1},
\label{K}
\end{equation}
where $\alpha$ is the strength of the turbulence. We also assume the factor $s_{\rm min}$ is given by \citep{kan15b}:
\begin{equation}
s_{\rm min}=\dfrac{1}{1+0.04K}.
\label{smin}
\end{equation}
We note that the back reaction from pebbles onto gas is not considered here for simplicity (see Appendix \ref{etazero}).

The ratio of the pressure gradient to the gravity of the central star is then,
\begin{equation}
\eta=-\dfrac{1}{2}\left(\dfrac{H_{\mathrm{g}}}{r}\right)^{2}\dfrac{\partial \ln{\rho_{\mathrm{g}}c_{\mathrm{s}}^{2}}}{\partial \ln{r}},
\label{eta}
\end{equation}
where $\rho_{\rm g}=\Sigma_{\rm g,local}/(\sqrt{2\pi}H_{\rm g})$ is the (local) gas density on the mid-plane.

\subsubsection{Migration of an embedded planet}
\label{migration}
The embedded planet migrates inward by Type I migration since we assume its mass is $20~M_{\rm E}$ (Earth mass) and do not include mass growth (planet growth will be covered in Paper II). The migration timescale is then given by \citep{tan02,ida08a}
\begin{equation}
\tau_{\rm mig}=\dfrac{1}{2.728+1.082p}\left(\dfrac{c_{\rm s}}{r_{\rm pl}\Omega_{\rm K,pl}}\right)^{2}\dfrac{M_{*}}{M_{\rm pl}}\dfrac{M_{*}}{r_{\rm pl}^{2}\Sigma_{\rm g,unp}}\Omega_{\rm K,pl}^{-1}.
\label{tmig}
\end{equation}

\subsection{Lagrangian Particle Model}
We use a Lagrangian particle model to calculate the distribution of pebbles in the discs. We consider super-particles of pebbles (i.e. groups of pebbles) and calculate their properties at each time step. We also calculate the properties of the gas disc at each time step and at the location of each of the super-particles by the equations in the above sections. We fix the Stokes number (the stopping time normalized by the Kepler time) of pebbles as ${\rm St}=0.1$. We insert $n_{\rm insert}$ super-particles at the outer boundary of the calculation area, $r_{\rm out}=50~{\rm au}$, at every time step with the mass of the super-particles as $m_{\rm peb,sp}=\dot{M}_{\rm peb}\Delta t/n_{\rm insert}$, where $\dot{M}_{\rm peb}$ and $\Delta t$ are the pebble mass flux and the duration of each time step, respectively. The particles are inserted  from $r_{\rm out}$ to $r_{\rm out}+v_{\rm drift}(t, r_{\rm out})\Delta t$ at intervals of $|v_{\rm drift}(t, r_{\rm out})|\Delta t/n_{\rm insert}$, where $v_{\rm drift}$ is the radial drift velocity of particles (Eq. (\ref{vdrift})). We set the duration of time step as $\Delta t=1/\Omega_{\rm K,pl}$, the inverse of the Keplerian frequency at the orbit of the planet, which becomes shorter as the planet migrates inward. When $\Delta t$ is relatively long we use a large value of $n_{\rm insert}$, whereas when $\Delta t$ is relatively short we fix $n_{\rm insert}$ at unity and change the probability of inserting a new particle, (i.e., a new particle is inserted at every times step), so that one new particle is inserted every ten years statistically. Note that we verify that this setup does not violate the mass conservation and the constant mass inflow flux of the solid particles in the calculation area (see Figure \ref{fig:totalmass} and its explanations in Section \ref{fiducial}). We treat the pebble mass flux as a parameter with $\dot{M}_{\rm peb}=10^{-5}$, $10^{-4}$, and $10^{-3}~M_{\rm E}~{\rm year}^{-1}$. The second value is almost the same as the one in \citet{lam14}, which is a typical value of the pebble flux in protoplanetary discs with typical properties. With this pebble flux, it takes more than $1~{\rm Myr}$ to form a first-generation planet of $20~M_{\rm E}$ by pebble accretion \citep{lam14}, which implies that our new proposed mechanism would happen at later times in the disc evolution. However, the pebble flux should depend on the metallicity of the discs and the first value can also be encountered \citep{bit18b}. Moreover, recent observations show that even a young disc ($<0.5~{\rm Myr}$ old) has gap structures, which suggests planets can exist much earlier than we predict by current planet formation models \citep{seg20}. In this paper, our focus is to propose a new possible scenario for planetesimal formation. More realistic models (e.g. including the initial growth of the first-generation planet, disc evolution, etc.) will be investigated in Paper II. The super-particles then drift inward until they get close to the orbit of the planet. The super-particles that passed through the planet's orbit or go to the region where $r>r_{\rm out}$ are removed in the next time step. The calculation stops when the planet reaches $r_{\rm in}=0.5~{\rm au}$. Sometimes, because of random motion induced by turbulence, some super-particles pass other other innermost super-particles. At each time step, we sort the super-particles radially using heap sort in order to compute the surface density of pebbles (see below).

\subsection{Radial transfer of pebbles}
\label{radial}
The speed of the radial motion of pebbles is determined by their radial drift and the effect of diffusion. The radial motion velocity is the sum of the two components:
\begin{equation}
v_{\rm r}=v_{\rm drift}+v_{\rm diff}.
\label{vr}
\end{equation}
Pebbles drift inward because they lose their angular momentum by suffering head wind from the gas disc rotating with sub-Kepler speed (due to the gas pressure gradient). The radial drift velocity is then \citep{whi72,ada76,wei77},
\begin{equation}
v_{\rm drift}=-2\dfrac{\rm St}{{\rm St}^{2}+1}\eta v_{\rm K},
\label{vdrift}
\end{equation}
where $v_{\rm k}=r\Omega_{\rm k}$ is the Kepler velocity.

In addition to radial drift, particles experience  a diffusion process that we model by a random walk \citep{cha11}. The radial diffusion velocity is given by:
\begin{equation}
v_{\rm diff}= \sqrt{\dfrac{2D}{\Delta t}}W_{\rm G}+\dfrac{D}{\rho_{\rm g,mid}}\dfrac{\partial \rho_{\rm g,mid}}{\partial r}
\label{vdiff}
\end{equation}
where $W_{\rm G}$ is a Gaussian random variable with a mean equal to zero and a variance $\sigma^{2}$ equal to unity. The first term of the right side of the equation represents the random walk and the second one is the systematic correction term arising from non-homogeneous diffusion. In reality, the direction of the particle motion changes when a particle collides with  another one, which occurs at time intervals equal to the correlation timescale $\tau_{\rm cor}\sim1/\Omega_{\rm K}$ \citep{you07}. In our calculations, the duration of the time step is $\Delta t=1/\Omega_{\rm K, pl}$, which is always shorter than the correlation timescale of each super-particle because the planetary orbit is located further in compared to the particles (particles inside the planetary orbit are removed from the calculation, see previous section). Numerically, at each time step, we calculate a new value for the first term (the stochastic one) with a probability equal to $\Delta t/\tau_{\rm cor}$. If a new value is not computed, we use the same value of the first term as in the previous time step. On the other hand,  the second term (the systematic one) is computed at every time step. This numerical procedure is the same as in \citet{cha11}. The turbulent diffusivity of the pebble particles, the viscosity, and the Schmidt number are respectively given by $D=\nu/{\rm Sc}$, $\nu=\alpha c_{\rm s}H_{\rm g}$ and ${\rm Sc}=(1+{\rm St}^{2})^{2}/(1+4{\rm St}^{2})$ \citep{sha73,you07}. We  include the diffusion effect only inside $40~{\rm au}$ to keep the incoming mass flux constant.

\subsection{Planetesimal Formation}
\label{planetesimals}
Many previous works found that planetesimals can form by streaming instability where pebbles (dust particles) can accumulate. Here, we define the condition for the planetesimal formation as $\rho_{\rm peb, mid}/\rho_{\rm peb, gas}>1$, the pebble-to-gas density ratio on the mid-plane is larger than unity \citep{you05,joh07,dra14}. In order to calculate this density ratio, we first compute the pebble surface density using the Lagrangian particles. We define the smoothed pebble surface density at radius $r_{i}$, as
\begin{equation}
\Sigma_{\rm peb}(r_{i})\equiv\dfrac{1}{2\pi r_{i}}\sum_{j\in{\rm support}}m_{{\rm peb,sp},j}W(|r_{i}-r_{j}|, h_{i}),
\label{Sigmap}
\end{equation}
where the weighting kernel $W$ is a Gaussian function of characteristic length $h_i$: 
\begin{equation}
W(|r_{i}-r_{j}|, h_{i})\equiv\dfrac{1}{\sqrt{\pi}h_{i}}\exp\left[-\left(\dfrac{r_{i}-r_{j}}{h_{i}}\right)^{2}\right],
\label{W}
\end{equation}
The sum is performed over all ''supporting particles'', which are  particles whose distance from  $r_i$ is shorter than $3h_{\rm i}$. The length scale $h_{i}$ is equal to the pebble scale height, $H_{\rm peb}$, at $r_{i}$. This is motivated by the fact that the size of the pebbles clumps which are converted to planetesimals is about the pebble scale height \citep{dra14}. The pebble scale height is given by \citep{you07},
\begin{equation}
H_{\rm{peb}}=H_{\rm{g}}\left(1+\dfrac{{\rm St}}{\alpha}\dfrac{1+2{\rm St}}{1+{\rm St}}\right)^{-1/2},
\label{Hpeb}
\end{equation}
and the (smoothed) mid-plane density of pebbles is then $\rho_{\rm peb,mid}=\Sigma_{\rm peb}/(\sqrt{2\pi}H_{\rm peb})$. Note that the smoothed value is only used for to evaluate whether the condition for planetesimal formation is fulfilled, and that the radial motion of the super-particles is independent from this smoothed surface density.

If the condition is satisfied, we insert a new planetesimal super-particle at the point of the pebble super-particle. The mass of the newly formed planetesimal super-particle, $m_{\rm pls,sp}$, is
\begin{equation}
m_{\rm pls,sp}=x_{\rm SI}m_{\rm peb,sp},
\label{mplssp}
\end{equation}
where $m_{\rm peb,sp}$ is the mass of the pebble super-particle. We then subtract the planetesimal super-particle mass from the pebble super-particle mass. The parameter $x_{\rm SI}$ is defined by the efficiency parameter $\epsilon_{\rm SI}$, the timescale of planetesimal formation by streaming instability $\tau_{\rm SI}$, and the calculation time step $\Delta t$ as $x_{\rm SI}\equiv\epsilon_{\rm SI}\Delta t/\tau_{\rm SI}$. We take $\tau_{\rm SI}=10~{\rm year}$ as the fiducial value and change it to $\tau_{\rm SI}=100$ and $1000~{\rm year}$ \citep{you05,joh07}. The parameter $\epsilon_{\rm SI}$ is considered as a free parameter in some previous works and we take $\epsilon_{\rm SI}=0.1$ \citep{dra14,sch18}. We check the importance of the parameter $\tau_{\rm SI}$ in Section \ref{tSI} and show that its value does not change the results significantly.

The mass and orbital positions of the planetesimal super-particles are fixed. However, we take the following steps every 1000 time steps. First we sort the planetesimal super-particles by heap sort. We then merge them if the super-particles are too close. We check the distance between the adjacent super-particles of planetesimals ($i$ and $i+1$) and keep it as
\begin{equation}
\Delta r_{{\rm pls}i,i+1}>\dfrac{r_{{\rm pls},i}+r_{{\rm pls},i+1}}{2}(10^{{\rm log}_{\rm 10}(r_{\rm out}/r_{\rm in})/1000}-1),
\end{equation}
where $r_{\rm in}$ and $r_{\rm out}$ are the inner and outer boundaries of the numerical domain. We finally calculate the planetesimal surface density by using similar equations as for pebbles (Eqs.(\ref{Sigmap}) and (\ref{W})). However, we assume the characteristic length for planetesimals as
\begin{equation}
1/ h_{{\rm pls},i}\equiv\dfrac{1}{2} (1/r_{{\rm pls,in},i}+1/r_{{\rm pls,out},i}),
\label{hplsi}
\end{equation}
where $r_{{\rm pls,in},i}\equiv\min(r_{{\rm pls},i+3}, r_{{\rm pls},n_{\rm pls}})$, $r_{{\rm pls,out},i}\equiv\max(r_{{\rm pls},i-3},r_{{\rm pls},1})$, and $n_{\rm pls}$ is the number of the planetesimal super-particles at each time.
\section{Results}
\subsection{Reference cases}
\label{fiducial}
We calculated the planetesimal formation in the reference cases and found that planetesimals form in wide areas of protoplanetary discs. The calculations were carried out with the conditions of Discs A and B, $\alpha=10^{-3}$, the half and normal speed of Type I migration, and $\dot{M}_{\rm peb}=10^{-5}$, $10^{-4}$, and $10^{-3}~M_{\rm E}~{\rm year}^{-1}$.

\subsubsection{Planetesimal formation from accumulated pebbles}
\label{accumulation}
\begin{figure}[t]
\centering
\includegraphics[width=0.8\linewidth]{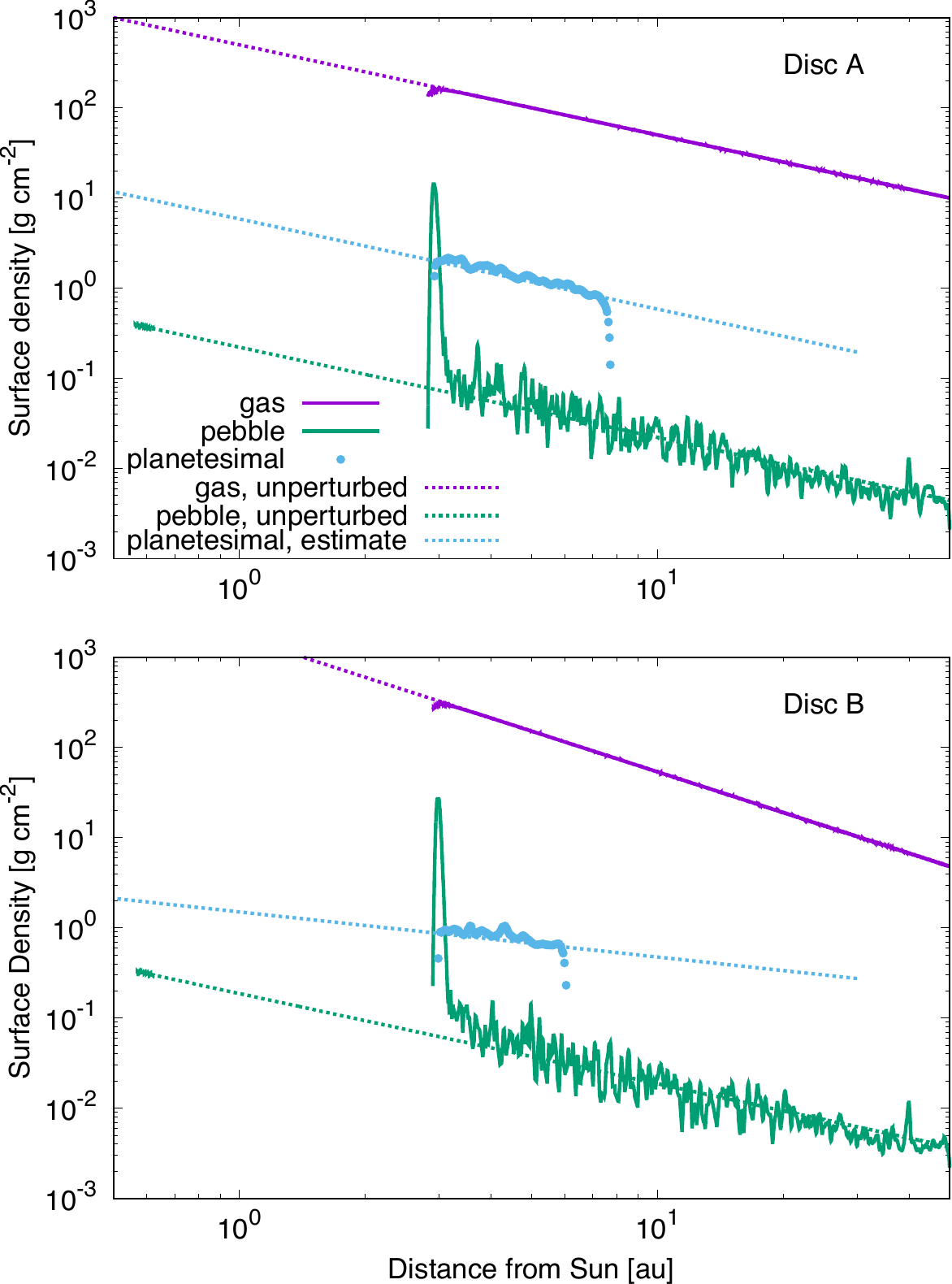}
\caption{Gas, pebble, and planetesimal surface density. The purple, green, and light blue solid plots represent these profiles, respectively, with Discs A (upper panel) and B (lower panel) at $t=3.8\times10^{5}~{\rm year}$. The pebble flux is $\dot{M}_{\rm peb}=10^{-4}~M_{\rm E}~{\rm year}^{-1}$ and the migration speed is the normal one. The purple, green, and light blue dotted lines represent the unperturbed gas surface density (Eq.(\ref{pebg})), the unperturbed pebble surface density (Eq.(\ref{gasg}0), and the estimated planetesimal surface density (Eq. (\ref{Sigmapl_est})), respectively.}
\label{fig:surface}
\end{figure}

First, we explain how to form planetesimals in our scenario using the results of the calculations with the  typical pebble flux ($\dot{M}_{\rm peb}=10^{-4}~M_{\rm E}~{\rm year}^{-1}$) and the normal migration speed. Figure \ref{fig:surface} represents the snapshots of the gas, pebble, and planetesimal surface density.\footnote{Note that we calculate the super-particles only outside the orbit of the planet so the profiles inside the orbit are not shown in the figure.} The gas surface density is carved by the influence of the planet and becomes lower than the unperturbed gas surface density (solid versus dotted purple, Eq.(\ref{gasg})) around the planet. The resulting bump in the gas pressure stops the inward drift of pebbles and the pebbles accumulate at the gas pressure maximum and make a pebble annulus (see Appendix \ref{etazero}). The pebble surface density dramatically increases there compared to the unperturbed pebble surface density, shown as dotted green line and computed using the following equation from the conservation of mass of pebbles,
\begin{equation}
\Sigma_{\rm peb,unp}=\dfrac{\dot{M}_{\rm peb}}{2\pi rv_{\rm drift,unp}},
\label{pebg}
\end{equation}
where $v_{\rm drift,unp}$ is the drift velocity (i.e. not including the diffusion effect) in the unperturbed disc (i.e. without the effect due to the planet). Due to the diffusion effect, the pebble surface density does not diverge to infinity and the annulus has some width (see Appendix \ref{validity}). The noise which exists the whole calculation region is also caused by the diffusion effect. Note again that this pebble surface density has been smoothed during its derivation with the pebble scale height (see Eq.(\ref{Sigmap})).

The accumulated pebbles are then converted to planetesimals by streaming instability, as the condition for planetesimal formation ($\rho_{\rm peb, mid}/\rho_{\rm peb, gas}>1$) is fulfilled (see Figure \ref{fig:rhopg}). The formation of planetesimals continues as  the planet migrates inwards and consequently the region of the planetesimal formation spreads inward. However, simulations show that planetesimals only form inside roughly $6-8~{\rm au}$. This value is similar to the one where the planet reaches its pebble isolation mass, which is roughly estimated by $M_{\rm iso}\approx20(r/(5~{\rm au}))^{3/4}~M_{\rm E}$ \citep{lam14}. The similarity between these two locations (outermost location of planetesimals and location where the planet reaches the isolation mass) is easy to understand: the pebble isolation mass is the mass beyond which the planet can not accrete pebbles anymore due to the gas pressure bump formed by the planet itself. These pebbles accumulating at the bump are converted to planetesimals, which is precisely the mechanism of planetesimal formation investigated in this paper. The final surface density of the planetesimals can be estimated by Eq.(\ref{Sigmapl_est}) (light blue dotted), as is explained in the next section.

\subsubsection{Characteristics of planetesimals}
\label{distribution}
\begin{figure}[t]
\centering
\includegraphics[width=0.8\linewidth]{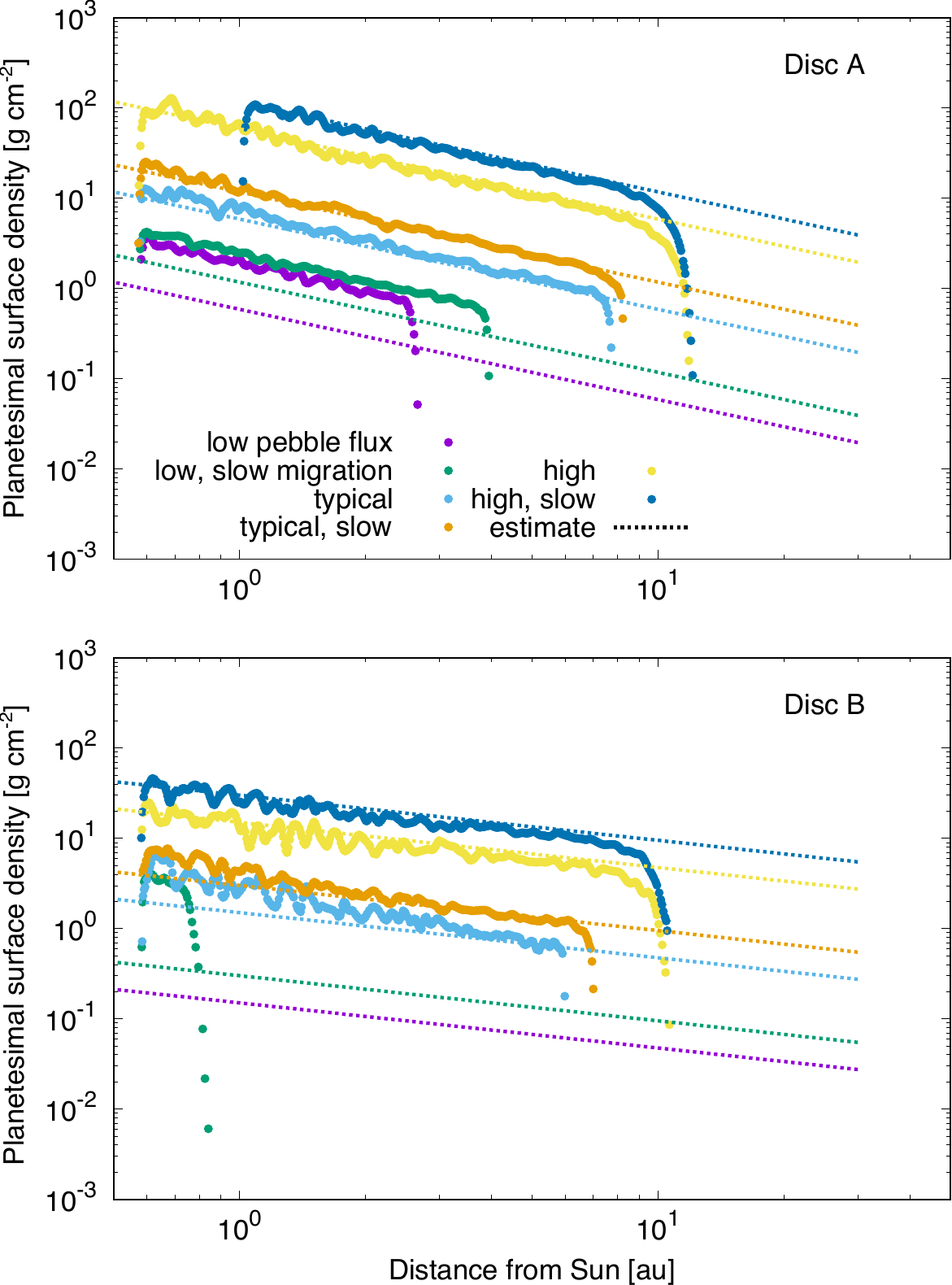}
\caption{Surface density profiles of planetesimals with Discs A (upper panel) and B (lower panel). The light blue and orange plots represent the planetesimal surface density with the normal and half (noted as `slow migration') speed of migration, respectively, with the pebble flux of $\dot{M}_{\rm peb}=10^{-4}~M_{\rm E}~{\rm year}^{-1}$ (typical case). The purple and green, and the yellow and blue curves are those with $\dot{M}_{\rm peb}=10^{-5}$ (low pebble flux) and $10^{-3}~M_{\rm E}~{\rm year}^{-1}$ (high). The dotted lines represent the estimated equivalents (Eq. (\ref{Sigmapl_est})). In Disc A with the high pebble flux and the half migration speed (blue), we stop the calculation when the planet reaches about $0.9~{\rm au}$ to save computer time. In Disc B, no planetesimals form with the low pebble flux cases.}
\label{fig:sigmap}
\end{figure}

\begin{figure}[t]
\centering
\includegraphics[width=0.8\linewidth]{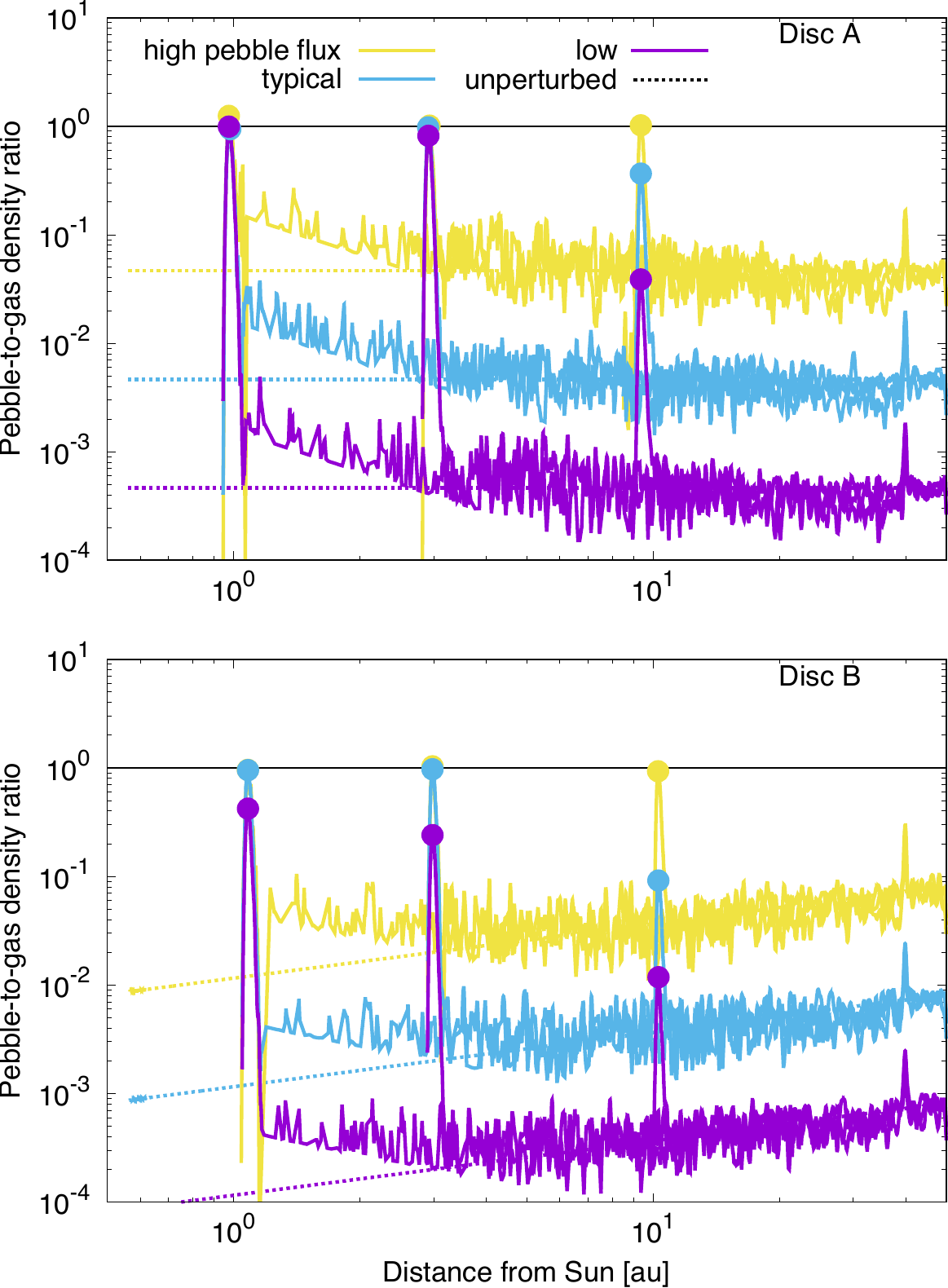}
\caption{Pebble-to-gas density ratios on the mid-plane with Discs A (upper panel) and B (lower panel). The yellow, light blue, and purple curves represent the snapshots of the density ratios with $\dot{M}_{\rm peb}=10^{-3}$, $10^{-4}$, and $10^{-5}~M_{\rm E}~{\rm year}^{-1}$, respectively, with the normal migration speed (i.e. the color variations are the same with those of Figure \ref{fig:sigmap}). The profiles are the snapshots at $t=3.0\times10^{5}$, $3.8\times10^{5}$, and $4.0\times10^{5}~{\rm year}$ (from right to left) for Disc A and $t=3.2\times10^{5}$, $3.8\times10^{5}$, and $3.9\times10^{5}~{\rm year}$ for Disc B. The circles are the maximum value for each of the profiles. The dotted lines are the unperturbed pebble-to-gas density ratios on the mid-planes (Eq. (\ref{rhopg})). The black lines represent where the ratio is unity.}
\label{fig:rhopg}
\end{figure}

We turn now to the characteristics of the planetesimals considering the other values of the pebble flux and migration speed. Figure \ref{fig:sigmap} shows that planetesimals form in wide regions of the discs except for the cases with $\dot{M}_{\rm peb}=10^{-5}~M_{\rm E}~{\rm year}^{-1}$ in Disc B. The position where the planetesimals start to form depends on the place where the planet reaches the isolation mass, as explained aboce, but also on the pebble mass flux. This latter dependence results from the fact that the condition to trigger streaming instability depends on the amount of pebbles accumulated at a specific point, itself strongly depending on the pebble flux.  Indeed, Fig. \ref{fig:rhopg} represents the snapshots of the pebble-to-gas density ratios on the mid-plane at three different time. The ratios increase as the pebbles get close to the planet and reach a maximum at the point where $\eta=0$ in all cases (see Appendix \ref{etazero}). However, the pebble-to-gas density ratio depends on the pebble mass flux, and the ratio can not reach unity with the low pebble flux in Disc B, which is the condition of the occurrence of streaming instability (Section \ref{planetesimals}). The unperturbed density ratio, far from the planet, can be actually estimated by the following equation,
\begin{equation}
\left(\dfrac{\rho_{\rm peb,mid}}{\rho_{\rm g,mid}}\right)_{\rm unp} = \dfrac{\Sigma_{\rm peb,unp}}{\Sigma_{\rm g,unp}}\dfrac{H_{\rm g}}{H_{\rm peb}},
\label{rhopg}
\end{equation}
which is proportional to $\dot{M}_{\rm peb}$ (the dotted lines). We note that the increase of the ratio starting from about $3~{\rm au}$ is not physical but it does not affect the profiles of the planetesimals (see Appendix \ref{validity}). The small variations at $40~{\rm au}$ are also not physical, and appear only because the diffusion effect is calculated only inside this location (see previous sections).

\begin{figure*}[t]
\centering
\includegraphics[width=0.8\linewidth]{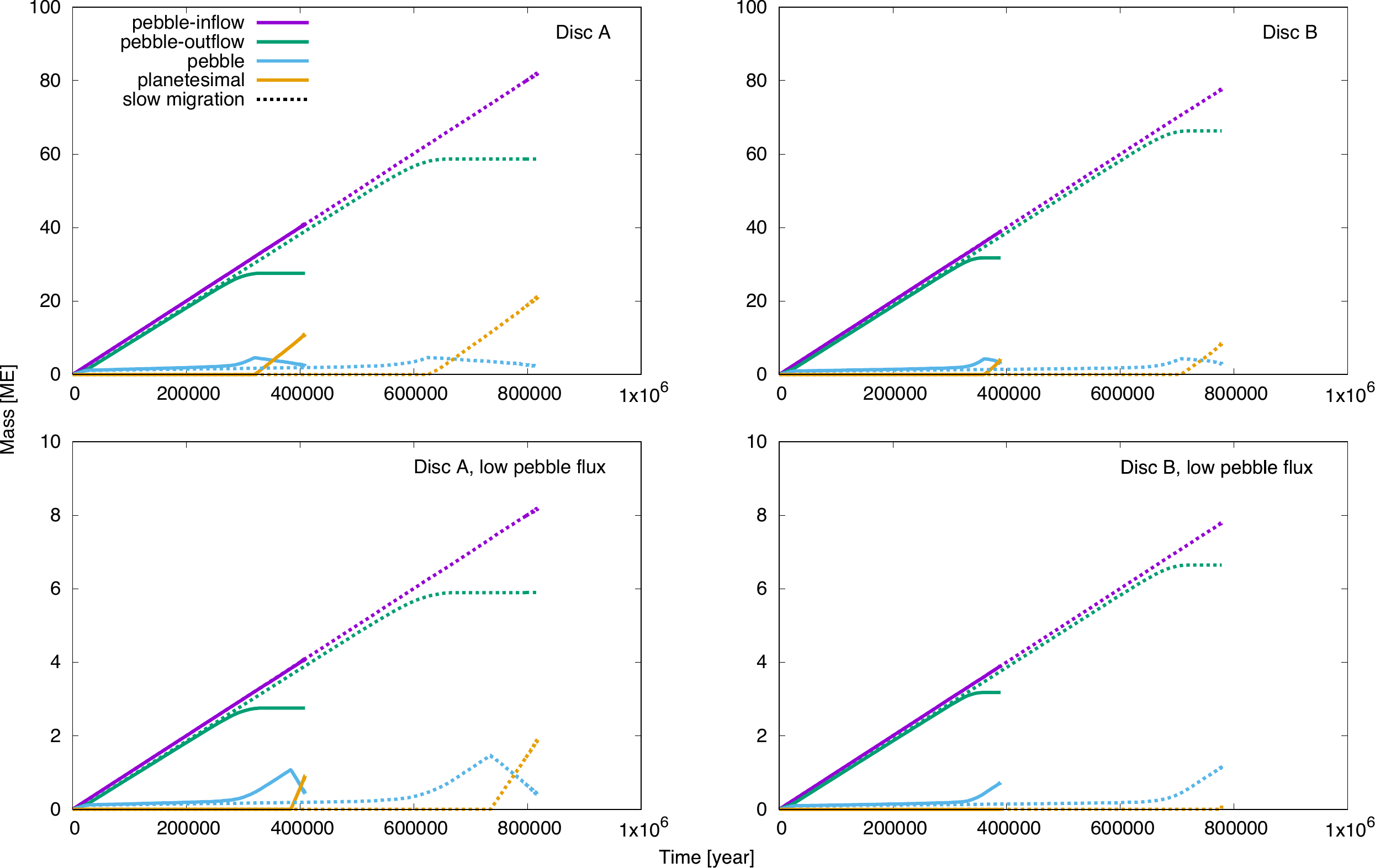}
\caption{Distribution of solid mass with Discs A and B (left and right panels, respectively). The pebble mass fluxes of the upper and lower panels are $\dot{M}_{\rm peb}=10^{-4}$ and $10^{-5}~M_{\rm E}~{\rm year}^{-1}$, respectively. The purple curves represent the cumulative mass of the pebbles flowing into the calculation area from the outside of $50~{\rm au}$. The green curves represent the cumulative mass of the pebbles flowing out from the calculation area through the planet's orbit. The light blue and orange curves are the total mass of the pebbles and planetesimals within the calculation area at each time, respectively. The dotted curves are those with the half migration speed.}
\label{fig:totalmass}
\end{figure*}

Figure \ref{fig:sigmap} shows that a simple equation can explain the planetesimal surface density very well once the planetesimals start to form. We found that the profiles are very well approximated by the following equation,
\begin{equation}
\Sigma_{\rm{pls, est}}\equiv\dfrac{\dot{M}_{\rm peb}}{2\pi rv_{\rm mig}},
\label{Sigmapl_est}
\end{equation}
where $v_{\rm mig}=r_{\rm pl}/\tau_{\rm mig}$ is (the absolute value of) the migration speed of the planet. This equation can be derived from the conservation of solid mass with the assumption that all of the pebbles approaching the planet are converted to planetesimals just after they start accumulating at the gas pressure bump. In other words, a kind of equilibrium or quasi-steady state between the flowing-in of pebbles and the formation of planetesimals must be present around the planet with inward migration for the formula to hold. Actually, the upper panels of Figure \ref{fig:totalmass} shows that the cumulative mass of the pebble flow into the calculation area from the outside of $50~{\rm au}$ (the purple curves) increases constantly, and the cumulative mass of the pebbles passing through the orbit of the planet (green) follows it, which means the pebbles steadily flow inward from the outer to inner regions passing through the planet. However, once the planetesimals start to form, the increase of the green curves becomes almost zero. The total mass of pebbles in the calculation area (light blue) is almost constant, which means the mass of accumulated pebbles dose not increase and they are converted to planetesimals immediately. Note that the difference of the two cumulative mass of the inflow and outflow pebbles (purple and green) is equal to the sum of the mass of the pebbles and planetesimals existing inside the calculation area (light blue and orange) in any cases. This shows the conservation of solid mass in the discs.

On the other hand, Figure \ref{fig:sigmap} also shows that the planetesimal surface density is larger than the estimate value in the case of $\dot{M}_{\rm peb}=10^{-5}~M_{\rm E}~{\rm year}^{-1}$ in Disc A. This is because the pebbles first accumulate at the gas pressure maximum and only after some time they start to be  converted into  planetesimals. The production rate of newly forming planetesimals is then larger than $\dot{M}_{\rm peb}$ because of the amount of pebbles that accumulated before the formation of planetesimals started. The left lower panel of Figure \ref{fig:totalmass} actually shows that the total mass of pebbles (light blue) decreases since that of the planetesimals (orange) starts increasing. This difference between the typical and high pebble flux on one side, and the low pebble fluxes on the other side results from the following two facts: 1) The time when the point where $\eta=0$ starts to exist is not different between the cases of the typical and low pebble fluxes because it is only determined by the gas and the planet in our calculation model. 2) However, the condition for planetesimal formation, $\rho_{\rm peb, mid}/\rho_{\rm peb, gas}>1$, is more difficult to fulfill in the low pebble flux case. As a result, the planetesimal surface density in the typical and high pebble flux cases is estimated correctly by Eq. (\ref{Sigmapl_est}) but in the case of the low pebble flux case there is some deviation because of the accumulated pebbles.

Figure \ref{fig:totalmass} also shows that the cumulative mass of the planetesimals in Disc A is larger than that in Disc B, and planetesimals do not form with the low pebble flux in Disc B. We discuss the total amount of the planetesimals also in Section \ref{comparison}.

\subsection{Parameter study}
\label{parameter}
We have investigated effect of different parameters of our model on the formation of planetesimal formation. We changed the strength of the turbulence, the mass of the protoplanetary discs, and the timescale of streaming instability.

\subsubsection{Strength of turbulence}
\label{turbulence}
\begin{figure}[t]
\centering
\includegraphics[width=0.8\linewidth]{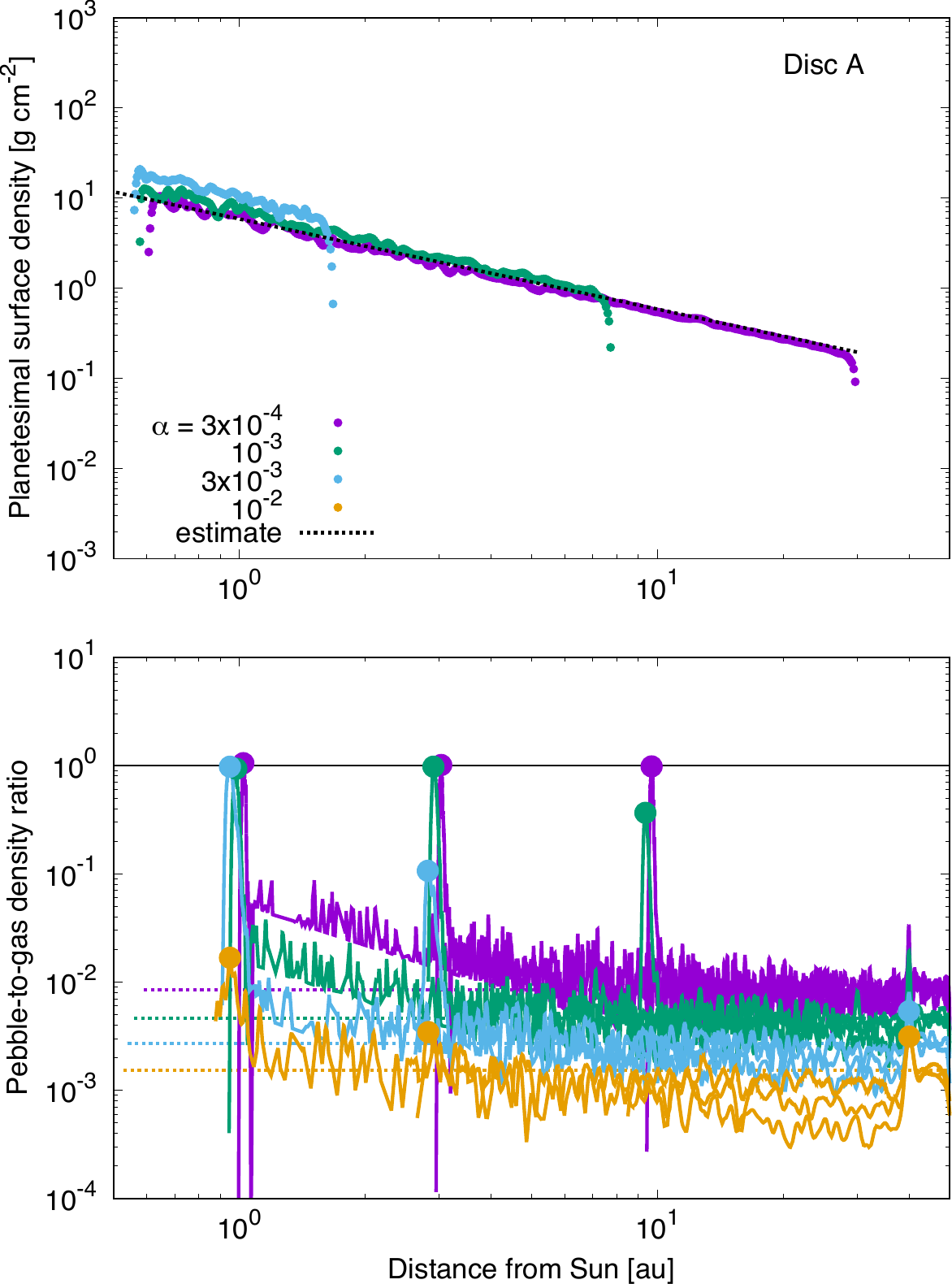}
\caption{Profiles with different strength of turbulence. The upper panel represents the planetesimal surface density with Disc A and $\dot{M}_{\rm peb}=10^{-4}~M_{\rm E}~{\rm year}^{-1}$. The lower panel is the snapshots of the pebble-to-gas density ratio on the mid-plane with the same condition at $t=3.0\times10^{5}$, $3.8\times10^{5}$, and $4.0\times10^{5}~{\rm year}$. The purple, green, light blue, and orange plots represent the profiles with $\alpha=3\times10^{-4}$, $10^{-3}$, $3\times10^{-3}$, and $10^{-2}$. The dotted line in the upper panel is the estimated planetesimal surface density by Eq. (\ref{Sigmapl_est}). No planetesimals form with $\alpha=10^{-2}$. The dotted lines in the lower panel are the unperturbed pebble-to-gas density ratios on the mid-planes (Eq. (\ref{rhopg})). The black line in the lower panel represents where the ratio is unity.}
\label{fig:alpha}
\end{figure}

Figure \ref{fig:alpha} represents the profiles of planetesimal surface density obtained with different values of the strength of turbulence. The upper panel shows that planetesimals form with $\alpha\leq3\times10^{-3}$ and the area where the planetesimals form is narrower as the turbulence is stronger. If the turbulence is too strong, $\alpha=10^{-2}$, no planetesimals form. These results can be explained by the facts that the vertical and horizontal diffusion of pebbles depend on the strength of turbulence. Equations (\ref{Hpeb}) and (\ref{rhopg}) show that the value of the unperturbed pebble-to-gas density ratio on the mid-plane is almost proportional to $1/\sqrt{\alpha}$, which is consistent with the profiles in the lower panel of Figure \ref{fig:alpha}. The figure also shows that as the turbulence becomes stronger, the distribution of the pebbles accumulating around the gas pressure maximum becomes wider due to the horizontal diffusion (see Eq.(\ref{vdiff})). These facts make it harder for the density ratio to reach unity, which is the condition for the occurrence of streaming instability. Figure \ref{fig:alpha} shows that the value of the planetesimal surface density does not depend on the strength of turbulence once they can form. This is consistent with the estimate by Eq. (\ref{Sigmapl_est}).

Figure \ref{fig:alpha} also shows that the point where the pebbles accumulate depends on the strength of turbulence. Strong turbulence, in other words strong diffusion, makes the gas pressure maximum closer to the planet because the gas structure is determined by the balance between the gas diffusion and the effect from the planet. In the case of $\alpha=10^{-2}$, the pebbles did not much accumulate at the gas pressure maximum due to the strong radial diffusion. We also note that the pebble-to-gas density ratios on the mid-planes are lower than the unperturbed ones around $20-40~{\rm au}$. The correlation timescale of the particles $t_{\rm corr}\equiv 1/\Omega_{\rm K}$ becomes longer as the distance from Sun becomes larger, resulting in the larger dispersion of the particles due to the rarely change of the direction of their motion (see Section \ref{radial}).

\subsubsection{Disc mass}
\label{discmass}
\begin{figure}[t]
\centering
\includegraphics[width=0.8\linewidth]{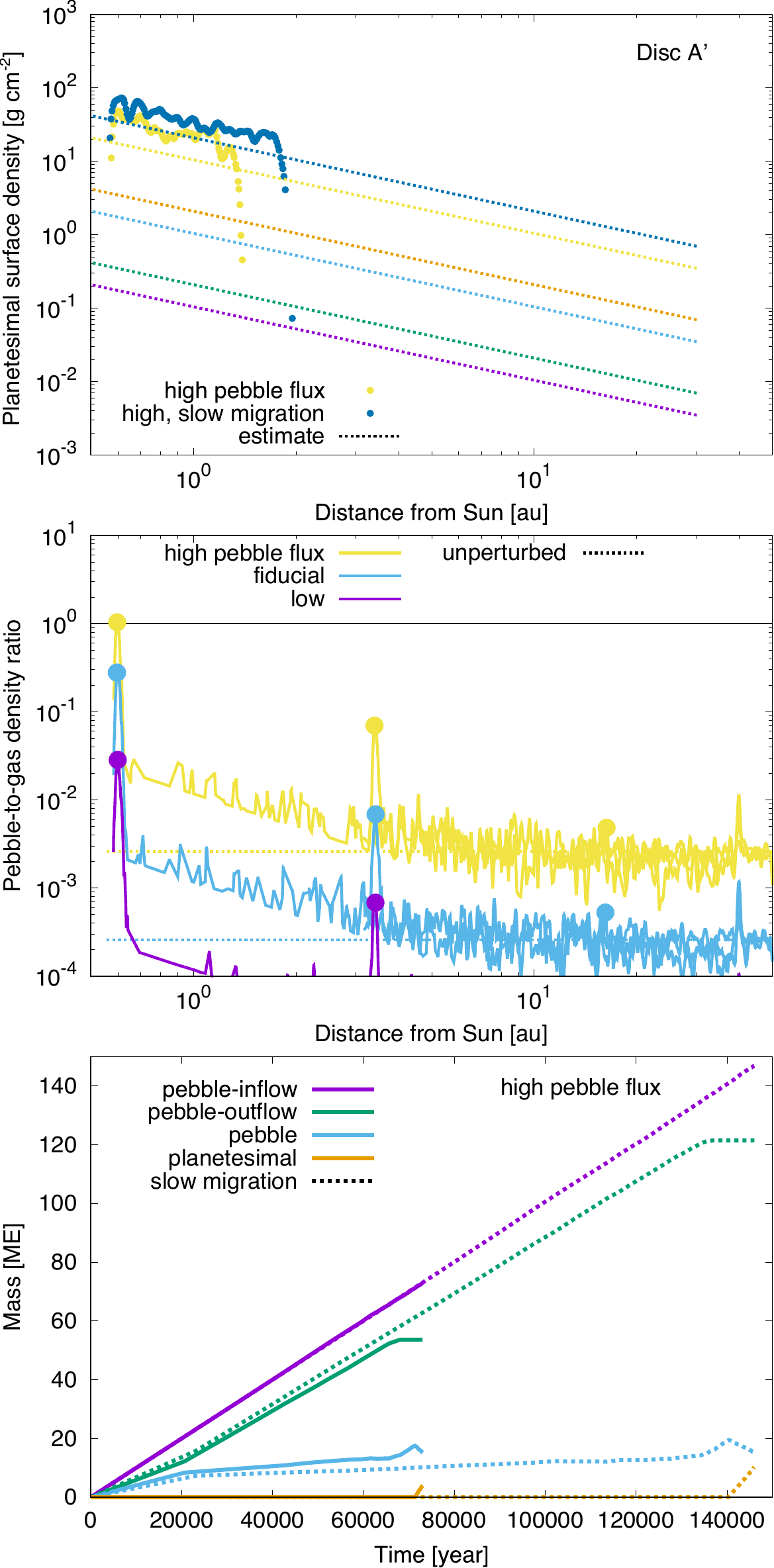}
\caption{The upper two and the bottom panels are same as Figures \ref{fig:alpha} and \ref{fig:totalmass}, respectively, but with a heavier gas disc, Disc A'. The yellow and blue plots in the upper two panels represent the profiles with the normal and half migration speed, respectively, with $\dot{M}_{\rm  peb}=10^{-3}~M_{\rm E}~{\rm year}^{-1}$. The orange plots are the profile with $\dot{M}_{\rm peb}=10^{-4}~M_{\rm E}~{\rm year}^{-1}$ and the normal migration speed. The light blue and purple curves in the middle panel are the profiles with $\dot{M}_{\rm peb}=10^{-4}$ and $10^{-5}~M_{\rm E}~{\rm year}^{-1}$, respectively, with the normal migration speed. In the bottom panel $\dot{M}_{\rm peb}=10^{-3}~M_{\rm E}~{\rm year}^{-1}$.}
\label{fig:discmass}
\end{figure}

We also investigated the dependency on the mass of the protoplanetary discs. Figure \ref{fig:discmass} shows that only in the cases with $\dot{M}_{\rm peb}=10^{-3}~M_{\rm E}~{\rm year}^{-1}$, planetesimals can form in Disc A', which is 10 times heavier than the reference case (Table \ref{tab:discs}). The middle panel explains this result: the pebble-to-gas density ratios on the mid-plane are lower than those in the reference (lower mass) cases and the ratio can reach unity only with the high pebble mass flux. Actually, Eq. (\ref{rhopg}) shows that the ratio (far from the planet) is proportional to the inverse of the gas surface density (and temperature) of the disc. This implies that the ratio of the pebble mass flux to the gas surface density is one of the important parameter for the condition of the occurrence of planetesimal formation. This parameter could be correspond to the metallicity of the protoplanetary discs, an effect that will be investigated in Paper II, including also the growth of pebbles from small dust particles. The top panel shows that the planetesimal surface density is higher than the value estimated by Eq. (\ref{Sigmapl_est}). The bottom panel shows that some pebbles accumulate at the gas pressure bump before the planet migrates, which makes a concentration of the formation of planetesimals at the inner region.

\subsubsection{Timescale of streaming instability}
\label{tSI}
\begin{figure}[t]
\centering
\includegraphics[width=0.8\linewidth]{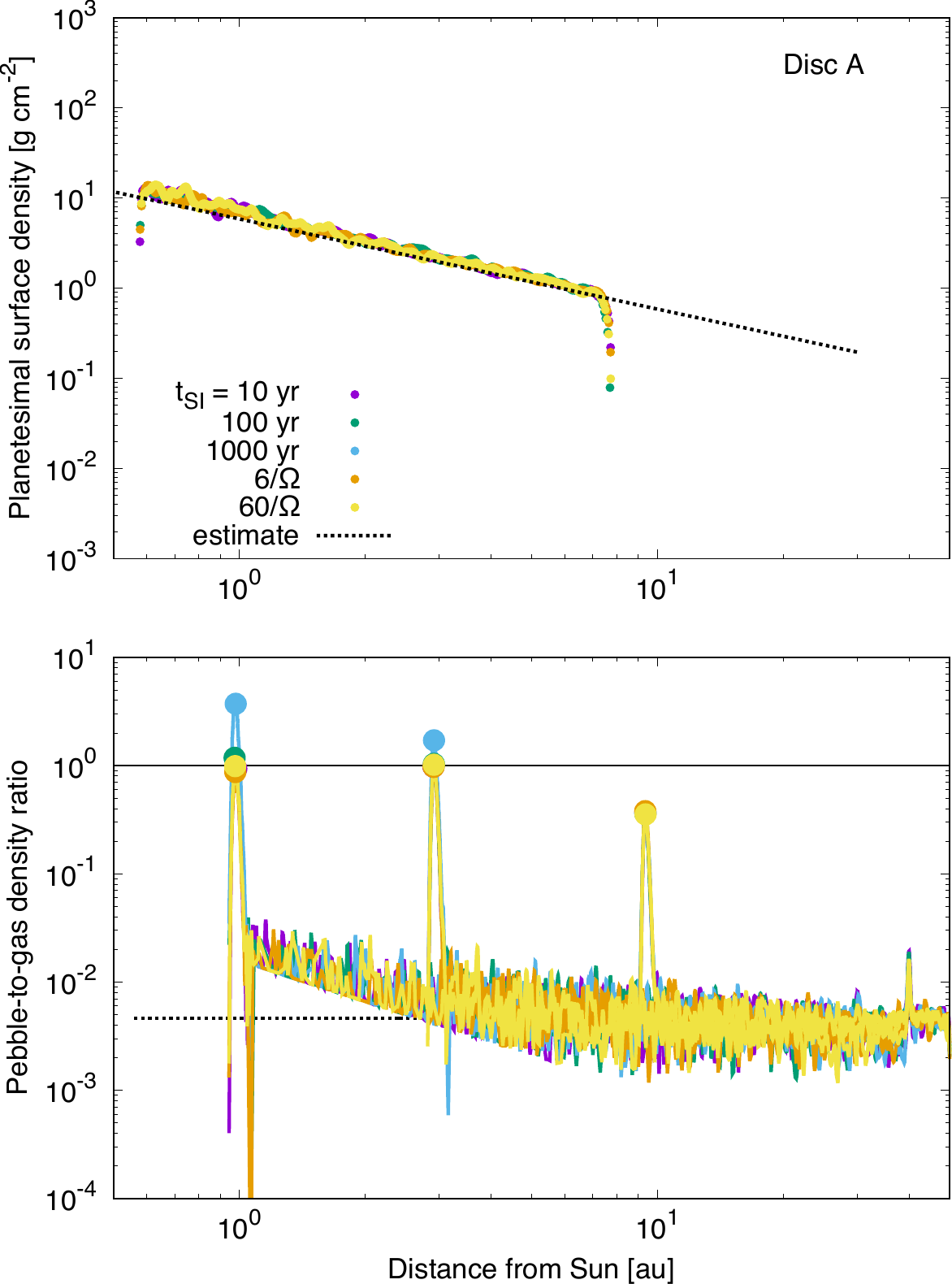}
\caption{Same as Figure \ref{fig:alpha} but with different timescale of streaming instability. The purple, green, light blue, orange, and yellow plots represent the profiles with $t_{\rm SI}=10$, $100$, $1000~{\rm year}$, and $6/\Omega_{\rm K}$ and $60/\Omega_{\rm K}~{\rm year}$, respectively, with $\dot{M}_{\rm peb}=10^{-4}~M_{\rm E}~{\rm year}^{-1}$ and the normal migration speed in Disc A.}
\label{fig:tSI}
\end{figure}

The timescale of planetesimal formation by streaming instability is still controversial. Previous works showed that the growth timescale of streaming instability, which is not necessarily the same with as the timescale of planetesimal formation by streaming instability itself, is about $10$ to $100~{\rm years}$ \citep{you05,joh07}. Moreover, the existence of the pebbles may change the gas structure and make $\eta$ smaller around the gas pressure maximum, which may make the growth timescale longer \citep{tak16}. We therefore checked the results assuming a longer formation timescale of streaming instability $\tau_{\rm SI}=100$ and $1000~{\rm years}$. We also investigated the cases where the streaming instability timescale depends on local the orbital time, with two different scalings: $\tau_{\rm SI}=6/\Omega_{\rm K}$ and $60/\Omega_{\rm K}$, according to the time between the instability is triggered and it reaches its non-linear saturation in previous simulation works \citep{joh07}. Figure \ref{fig:tSI} shows that the planetesimal surface density does not depend on the timescale significantly. If the timescale is long, a larger fraction of pebbles remain as pebbles at each time step even if the condition of the formation is satisfied (see Section \ref{planetesimals}). This is especially the case in the disc inner parts (light blue). These remaining pebbles, however, do not drift any more and stay at the gas pressure maximum, which will be converted to planetesimals in the subsequent time steps. The remaining pebbles can be seen in the lower panel of Figure \ref{fig:tSI} as the peaks of the profiles which have higher value than unity.

\section{Discussions}
\subsection{Comparison to population synthesis works}
\label{comparison}
There have been no observations of the planetesimal discs so we can not directly discuss the validity of our new mechanism of planetesimal formation. However, there have been many population synthesis works that almost reproduced the distributions of the exoplanets \citep[e.g.][]{ida04a,mor09a}.

The slopes of the profiles of the planetesimal surface density especially affects the subsequent planet formation. As the slope is steeper, the final distribution of the planetary orbits concentrate at the inner regions of the systems \citep{len19,voe20}. From Eqs. (\ref{tmig}) and (\ref{Sigmapl_est}), this important slope can be estimated simply as $p-q-1.5$, where $-p$ and $-q$ are the slopes of the gas surface density and the disc temperature. In the cases of Discs A and B, the slopes are then $-1$ and $-0.5$, respectively. Most of the previous population synthesis works considered the minimum-mass solar nebular model and that the slope was $-1.5$ under the assumption that dust particles are fully converted to planetesimals in-situ \citep[e.g.][]{ida04a,mor09a}. This assumed slope is steeper than that of the planetesimals formed by our now mechanism in Disc B. From observations the properties of the gas discs should be $p=1$ (if the dust-to-gas surface density ratio is uniform in the whole discs) and $q=0.5$, which are also the same with the assumptions of Disc A \citep{and10}. The slope of the planetesimals which form in Disc A is therefore consistent with the observations through the assumption that small solid particles are fully converted to planetesimal in-situ. We note that the distributions of the planets in population synthesis models may change because the region where planetesimals are formed  does not spread until the outermost part of the discs. We however note that the precise outer boundary of the planetesimal disc depends on a combination of the pebble flux and the place where the first-generation planet reaches the isolation mass. This in turn depends on the thermal structure of the protoplanetary disc, and will be the subject of more detail studies in Paper II.

The amount of formed planetesimals is also an important outcome of the model. In the cases of the typical pebble mass flux, $\dot{M}_{\rm peb}=10^{-4}~M_{\rm E}~{\rm year}^{-1}$, the total mass of the planetesimals is of the order of $11$ and $21~M_{\rm E}$ for the normal and half migration speed in Disc A, and is $4.2$ and $8.6~M_{\rm E}$ in Disc B, respectively (see Figure \ref{fig:totalmass}), which are smaller than the assumptions in population synthesis works. However, in the cases of 10 times higher pebble mass flux, $\dot{M}_{\rm peb}=10^{-3}~M_{\rm E}~{\rm year}^{-1}$, the total mass is about $127$ and (more than) $240~M_{\rm E}$ (Disc A) and $58$ and $117~M_{\rm E}$ (Disc B), respectively, which are more than 10 times larger than the typical flux cases. The reason is that the surface density of planetesimals is almost proportional to the pebble mass flux (see Eq. (\ref{Sigmapl_est})) but the outermost orbital position where  planetesimals start to form also depends on the pebble flux (see Figure \ref{Sigmap}). Both effects combine to give a total mass of formed planetesimals depending more than linearly on the pebble flux.

From the viewpoint of surface density, at $2.7~{\rm au}$, the planetesimal surface density is about $2.3$ and $4.7~{\rm g~cm^{-2}}$ for the normal and half migration speed in Disc A, and is $1.0$ and $2.1~{\rm g~cm^{-2}}$ in Disc B, respectively, which are also smaller than that of the MMSN model, about $6.8~{\rm g~cm^{-2}}$ \citep{hay81}. However, if the pebble mass flux is 10 times higher, the surface density is between $20$ and $43~{\rm g~cm^{-2}}$ (Disc A) and $7.2$ and $17~{\rm g~cm^{-2}}$ (Disc B), respectively, which are larger than that of the MMSN model. Therefore, the new model of planetesimal formation we mention here may lead to relatively high values of the planetesimal surface density.

In conclusion, the amount of planetesimals which is formed by our new mechanism depends strongly on the pebble mass flux and the migration speed of the embedded planet. We will show detailed discussion of the characteristics of the planetesimals in Paper II, in which the formation process will be calculated by the models including both of the mass growth of pebbles and the embedded planet. We emphasize finally the fact that planetesimals may form not only by the formation mechanism proposed in this paper, but also by other mechanisms that generally occur at specific locations. The resulting distribution of the planetesimals will therefore likely  be the sum of the results predicted by the different mechanisms.

\subsection{Effect of the planet mass}
\label{heavier}
\begin{figure}[t]
\centering
\includegraphics[width=0.8\linewidth]{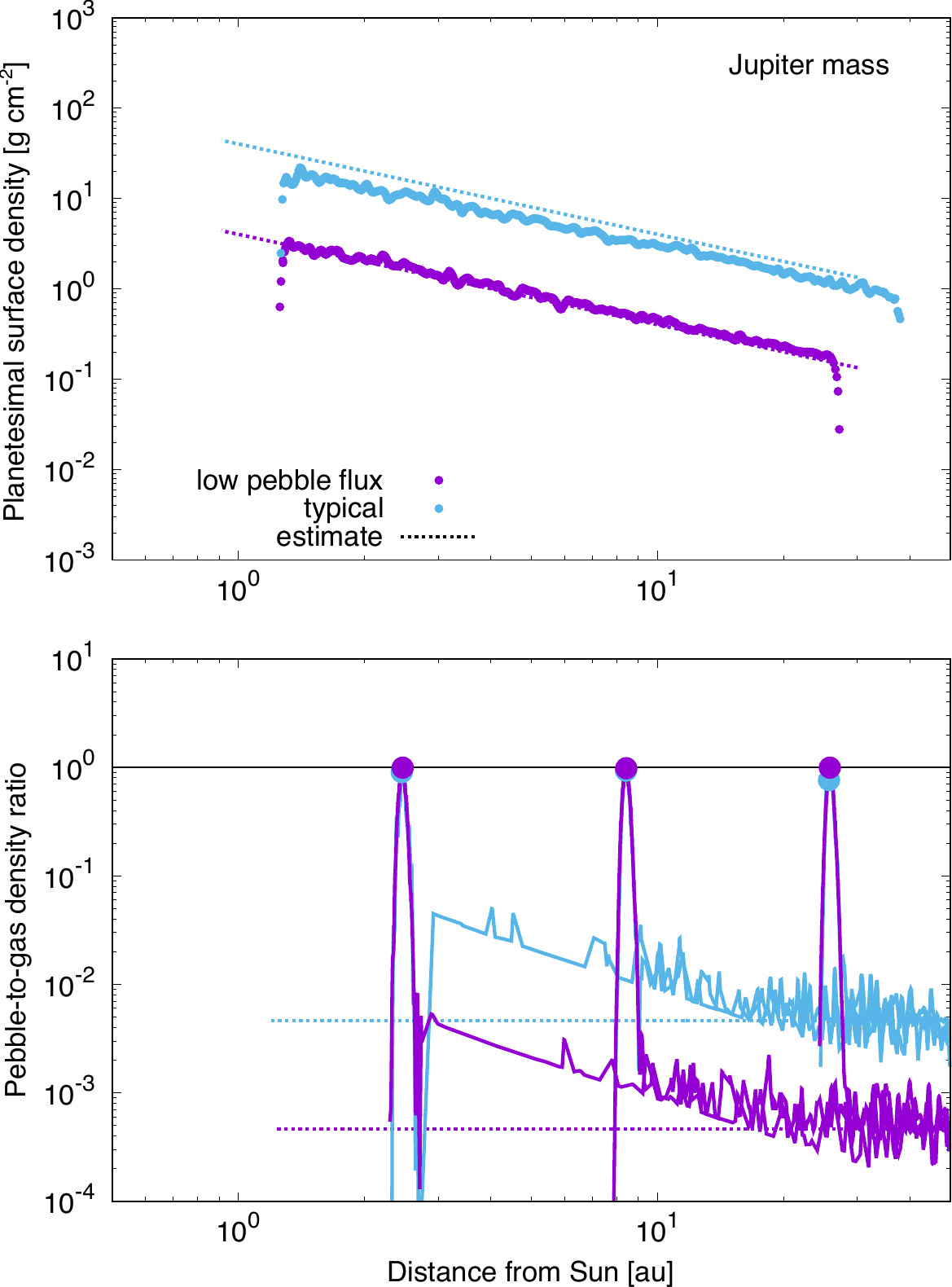}
\caption{Same as Figure \ref{fig:alpha} but with a Jupiter mass planet and its type II migration. The purple and light blue plots represent the profiles with $\dot{M}_{\rm peb}=10^{-5}$ and $10^{-4}~M_{\rm E}~{\rm year}^{-1}$ in Disc A, respectively. The rate to insert the super-particles of pebbles is $10$ times lower and the duration of the time step is $10$ times longer than those of the normal cases to make the calculation time shorter. We stop the calculations when the planets reach about $0.9~{\rm au}$. The time of the snapshots in the lower panel are $t=1.1\times10^{6}$, $2.3\times10^{6}$, and $2.7\times10^{6}~{\rm year}$.}
\label{fig:heavy}
\end{figure}

We have primarily investigated the planetesimal formation following a planet with $20~M_{\rm E}$ and Type I migration. However, if the growth of the planet is faster than its migration, the planet glow larger and the type of migration changes to Type II. We therefore checked the planetesimal formation with a planet with $1~M_{\rm J}$ and Type II migration. Here, we assume the migration timescale is the same as the advection speed of gas for simplicity, $\tau_{\rm migII}=2r^{2}/3\nu$. Figure \ref{fig:heavy} represents the results with these assumptions. \footnote{Note that we change some numerical parameters to reduce the calculation time (see the caption of the figure).} In the cases of a Jupiter mass planet, planetesimals start to form from the farther regions of the disc compared to the cases with a $20~M_{\rm E}$ planet and they form even in the case with $\dot{M}_{\rm peb}=10^{-5}~M_{\rm E}~{\rm year}^{-1}$. This is because the heavier planet carves the disc deeper and it makes the accumulation of pebbles easier (see Section \ref{gap}). The planet also reaches its pebble isolation mass from the start of the calculations, when the planet is at $r_{\rm pl}=30~{\rm au}$. Interestingly enough, it means that our mechanism could be extremely efficient at forming planetesimals following the formation of a first generation planet by disc instability. Indeed, disc instability models in general predict the formation of planets much more massive than the typical value of pebble isolation mass, close to the mass of Jupiter \citep[e.g.][]{may02} Furthermore, the planet formation by disc instability is so quick $(\sim10^{3}$ years) that enough amounts of pebbles are still drifting in the discs \citep{bit18b}.

The planetesimal surface density is $11~{\rm g~cm^{-2}}$ at $2.7~{\rm au}$ in Disc A with $\dot{M}_{\rm peb}=10^{-4}~M_{\rm E}~{\rm year}^{-1}$, which is higher than that of the reference case. The speed of Type II migration is much slower than that of Type I migration, resulting in the higher surface density, which is as well as explained by Eq. (\ref{Sigmapl_est}). Interestingly, the slopes of the profiles of the planetesimal surface density, which are $-1$ in Disc A, are the same with those with the planets migrating by Type I migration. This is because the $r$ dependency of the timescale of Type II migration is the same with that of Type I migration (Eq. (\ref{tmig})). In other words, the slopes of the planetesimal surface density formed by the mechanism we propose in this paper do not depend on the types of migration. 
 
\section{Conclusions}
We  propose in this paper a new mechanism of planetesimal formation which can operate in broad areas of protoplanetary discs. Once a first-generation planet, with a large enough mass, forms, planetesimals form by streaming instability from accumulating pebbles at the gas pressure maximum induced by the planet, and the planetesimal formation regions spreads inwards following the  migrating planet. We investigated this scenario by a 1D Lagrangian super-particle model with some different conditions of protoplanetary discs, pebbles, and migrating planets.

We then found that planetesimals can form with typical conditions and the results depend on some parameters, pebble mass flux, migration speed of the planet, surface density of the gas disc, and strength of turbulence. The total amount of the planetesimals strongly depends on the inflow of pebbles. The orbital position where planetesimals start to form is close to where the planet reaches is large enough to induce an accumulation of pebbles (the pebble isolation mass) but it also depends on the pebble mass flux. Once the planetesimals start to form, almost all pebbles drifting from the outer region are converted to the planetesimals as soon as they are trapped at the gas pressure bump. Therefore, the surface density of the planetesimals can be estimated from the conservation of solid mass by a very simple equation (Eq. (\ref{Sigmapl_est})), which is proportional to the pebble mass flux and inversely proportional to the migration speed of the planet. The slope of the profile of the planetesimal surface density is $p-q-1.5$, where $-p$ and $-q$ are the slopes of the gas surface density and the disc temperature. The slope  is determined by the $p$ and $q$ dependency of the migration speed. On the other hand, in the cases where the condition for planetesimal formation is fulfilled only close to the inner region of the disc (low pebble flux, heavier disc), the planetesimal surface density tends to be higher than the analytical estimate. This is because the pebbles accumulate at the gas pressure bump before they start to be converted into planetesimals. If the planet has grown enough for migrating by Type II, much slower migration, the surface density of the planetesimals can be larger and the radial distribution of the formed planetesimals is much broader than those with a Type I migrating planet. This means if the first-generation planet is formed by disc instability, a lot of planetesimals can form subsequently. On the other hand, the slopes of the planetesimal surface density in the cases of Type II migration are the same with those in the Type I migration cases. These results and the simple estimate ways could suggest the initial conditions of population synthesis works.

\begin{acknowledgements}
We thank the referee, Joanna Dr{\k{a}}{\.z}kowska, for the very useful comments. This work has been carried out within the frame of the National Centre for Competence in Research PlanetS supported by the Swiss National Science Foundation (SNSF). The authors acknowledge the financial support of the SNSF.
\end{acknowledgements}

\bibliographystyle{aa}
\bibliography{bib1}

\begin{appendix} 
\section{Radial positions of the accumulating pebbles}
\label{etazero}
\begin{figure}[t]
\centering
\includegraphics[width=0.8\linewidth]{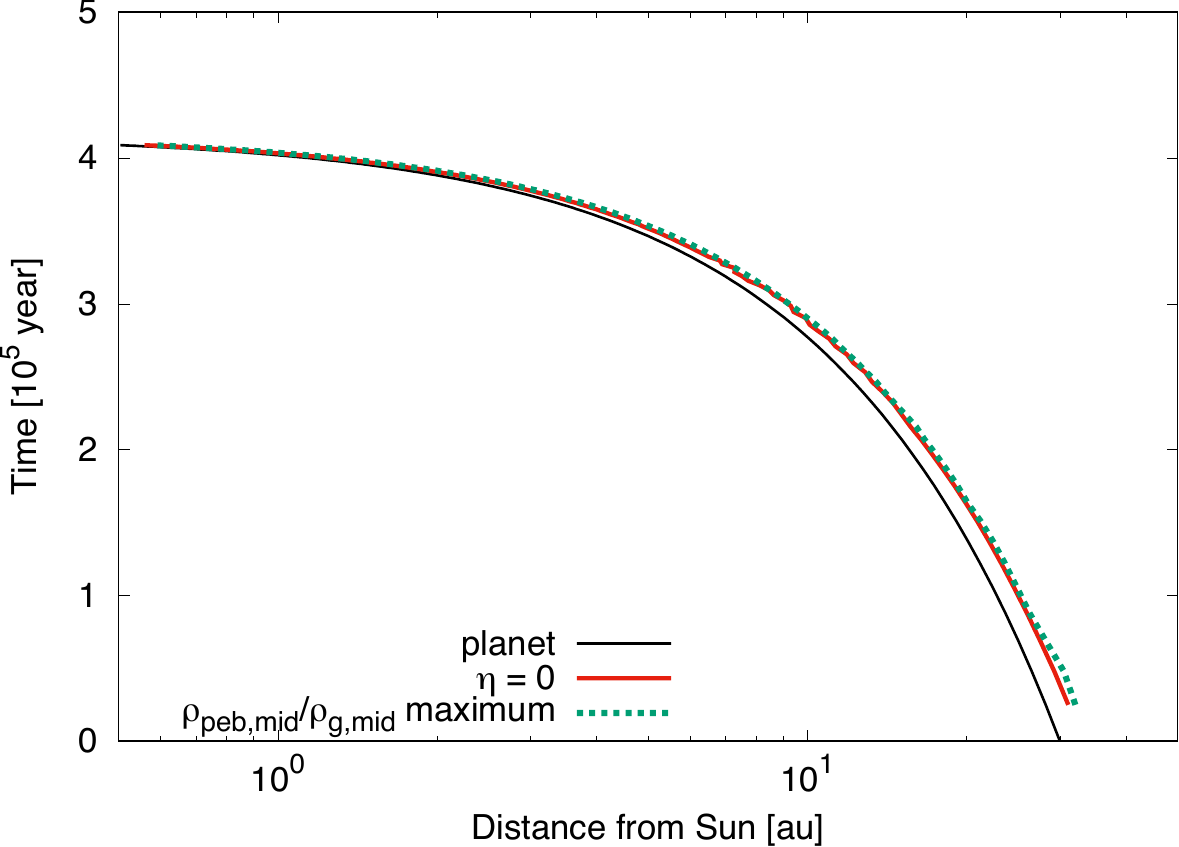}
\caption{Positions of the orbit of the planet, the gas pressure maximum, and the pebble-to-gas surface density ratio's maximum. Each of the colors is black, red, and dotted light green, respectively. We take the position of the gas pressure maximum as that of the pebble super-particle who has the smallest $|\eta|$ value. The conditions are $\dot{M}_{\rm peb}=10^{-4}~M_{\rm E}~{\rm year}^{-1}$, the normal migration speed, and Disc A.}
\label{fig:etazero}
\end{figure}

The pebbles accumulate at the gas pressure maximum, resulting in the formation of planetesimals by streaming instability there. Figure \ref{fig:etazero} showed that the radial positions of the gas pressure maximum and the pebble-to-gas density ratio's maximum is always at almost the same orbit. The accumulation point migrates inward as the gas pressure maximum migrates inward. The reason of the migration of the gas pressure maximum is, of course, that of the planet. The migration speed of the planetesimal formation point is not so different from that of the planet. As a result, the migration speed of the planet, $v_{\rm mig}$, can be used in Eq. (\ref{Sigmapl_est}) as the substitution of that of the position of the planetesimal formation, which makes the estimate equation simple.

We also show that the width of the annuli of the accumulating pebbles is consistent with an analytical prediction. Here we predict the width by comparing the two timescales, the diffusion and drift timescales of pebbles. The diffusion timescale is $t_{\rm diff}=\Delta r_{\rm diff}^{2}/(2D)$, where $\Delta r_{\rm diff}$ is the width of the pebble annulus. The drift timescale is $t_{\rm drift}=r/|v_{\rm drift,unp}|$, where we calculate the drift velocity ignoring the structure of the gas pressure bump. In reality, the velocity becomes slower as the pebbles approach the planet due to the decreasing $\eta$, so we can only predict the largest value of the width of the pebble annuli. Substituting those equations to $t_{\rm diff}=t_{\rm drift}$, we get the following estimate,
\begin{equation}
\Delta r_{\rm diff}\approx\left(\dfrac{\alpha}{\rm St}\dfrac{1}{\eta}\right)^{1/2}H_{\rm g},
\label{rdiff}
\end{equation}
and, for example, in the case of Disc A and $\alpha=10^{-3}$,
\begin{equation}
\Delta r_{\rm diff}\approx0.09\left(\dfrac{r}{\rm au}\right)~[{\rm au}].
\label{rdiffa}
\end{equation}
which is consistent with the obtained profiles of the accumulating pebbles in our calculations (see Figure \ref{fig:rhopg}).

We use some simplified formulas for the calculations of the gas structure around the embedded planet in this work (see Section \ref{planet}). In particular, the following effects which may change the gas and pebble radial distribution are not included; the back-reaction from the accumulating pebbles (dust) onto the gas, the effects of the spiral wakes which can be calculated only in 2-D simulations, and the effects of the waves induced by the migration of the planet. However, recent two-fluid (gas and dust) 2-D hydrodynamic simulations suggest that these effects can make the annuli of the accumulating pebbles wider but the pebble-to-gas density ratios are still high there \citep{kan18,dra19,surprep}.

\section{Validity of the calculations}
\label{validity}
\begin{figure}[t]
\centering
\includegraphics[width=0.8\linewidth]{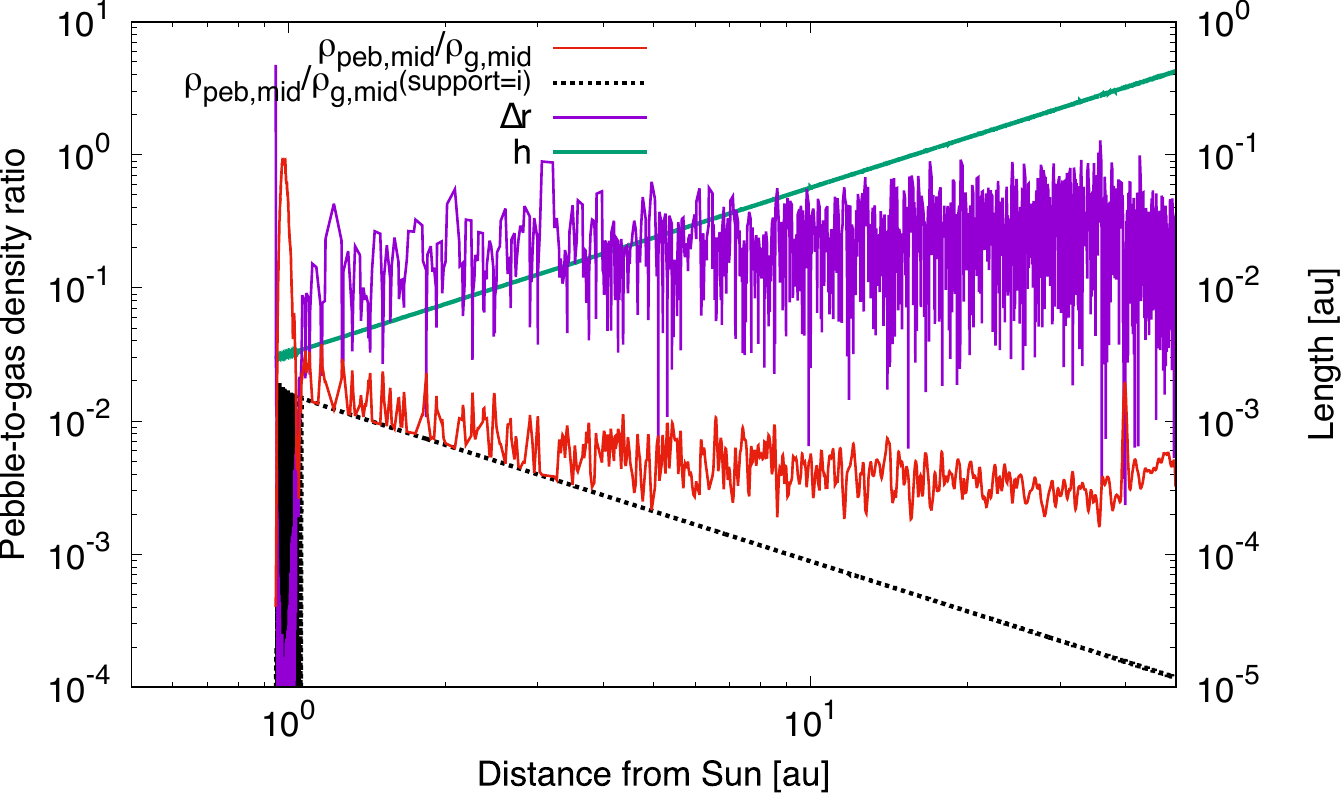}
\caption{Effect of the number of the super-particles. The curves are the snapshot of the profiles at $t=4.0\times10^{5}~{\rm year}$ with $\dot{M}_{\rm peb}=10^{-4}~M_{\rm E}~{\rm year}^{-1}$ and the normal migration speed in Disc A. The red, black dotted, purple, and green curves represent the smoothed pebble-to-gas density ratio on the mid-plane, the same ratio when the supporting particle is only the one at the orbit, the distance between the adjacent super-particles, and the characteristic length, respectively.}
\label{fig:validity}
\end{figure}

We investigate the validity of the calculations especially about the number of the super-particles of pebbles. The number larger, the more accurate results we can get but the calculations need much longer time as the number increases. The distance between the particles depends on the number of the particles, which should be shorter than the characteristic length at the orbit. We insert the super-particles with the rate that about one super-particle at one year. Figure \ref{fig:validity} shows that the characteristic length (green), which is equal to the pebble scale height, begins to be shorter than the distance between the adjacent super-particles (purple) around $5~{\rm au}$. The distance is wider than the characteristic length so unnatural structure arises on the profile. The pebble-to-gas density ratio on the mid-plane at $r_{i}$, $\rho_{\rm peb,mid}/\rho_{\rm g,mid}$, (red) is always larger than the density ratio of the case that the the supporting particle is only the particle $i$ (see Eqs. (\ref{Sigmap}) and (\ref{W})),
\begin{equation}
\left(\dfrac{\rho_{\rm peb,mid}}{\rho_{\rm g,mid}}\right)_{{\rm support}=i}\equiv\dfrac{m_{{\rm peb,sp},i}}{2\pi r_{i}}\dfrac{1}{\sqrt{\pi}h_{i}},
\label{base}
\end{equation}
which is plotted as a black dotted line in the figure. The total mass of each super particle is almost the same, $m_{\rm peb,sp}\approx\dot{M}_{\rm peb}\times1~{\rm year}$, so the profile is a straight line except where the particles are converted to the planetesimals. Therefore, the (smoothed) pebble-to-gas density ratio is not calculated correctly inside about $3~{\rm au}$. However, Figure \ref{fig:validity} also shows that the particles accumulate around the pressure bump and the distance between the particles is much shorter than the characteristic length there. The bottom value of the ratio is also much smaller than unity, the condition for planetesimal formation. The points are 1) the motion of the super-particles is independent from the properties of the other particles in our calculation model so the insufficient number of particles does not matter to the distribution of the particles, and 2) the density ratio is needed only when it reaches unity to get the correct profiles of the planetesimal surface density. Therefore, the calculations in this work are valid enough for the discussion of the planetesimal formation.

\end{appendix}
\end{document}